\documentclass[twocolumn,aps,prl,showpacs,superscriptaddress]{revtex4}
\usepackage{graphicx}
\usepackage{amsfonts}
\usepackage{amsmath}
\usepackage{amssymb}

\def\endproof{\vrule height6pt width6pt depth0pt}

\begin{document}

\title{Minimal proofs of state-independent contextuality}


\author{Ad\'an Cabello}
 \affiliation{Departamento de F\'{\i}sica Aplicada II, Universidad de
 Sevilla, E-41012 Sevilla, Spain}
 \affiliation{Department of Physics, Stockholm University, S-10691
 Stockholm, Sweden}


\date{\today}



\begin{abstract}
It has been recently shown that state-independent contextuality (SIC) is a fundamental resource linked with a type of nonlocality which cannot be improved
by nonsignaling resources. Therefore, it is of fundamental importance to identify the simplest sets of quantum observables needed to prove SIC. We show
that $d+10$ rank-1 projectors are sufficient to prove SIC for any physical system in dimension $d>2$. This result outperforms both the best Kochen-Specker proofs and the results presented by Yu and Oh in arXiv:1112.5513v1.
\end{abstract}



\maketitle


The predictions of quantum mechanics (QM) cannot be reproduced by noncontextual theories in which measurement results are independent of compatible measurements. This is known as quantum contextuality and was first proven by Specker \cite{Specker60}, Bell \cite{Bell66}, and Kochen-Specker (KS) \cite{KS67}. Quantum contextuality may be state-independent: for any given dimension $d\ge 3$, there are universal sets of observables which prove contextuality for any state of the system. It has been recently shown that state-independent contextuality (SIC) is a fundamental resource linked with a type of nonlocality which cannot be improved by nonsignaling resources \cite{Cabello11}. Therefore, it is of fundamental importance to identify the simplest sets of quantum observables needed to prove SIC.

All known proofs of SIC can be expressed in terms of rays in $\mathbb{C}^d$. There are two type of proofs of SIC: (i) {\em KS sets}, defined as those for which there is no map $f:\mathbb{C}^d \rightarrow \{0,1\}$ for $d\ge 3$ such that for all orthonormal bases $b \subseteq \mathbb{C}^d$, $\sum_{v \in b} f(v)=1$ \cite{KS67}. (ii) {\em SIC sets}, defined as those for which there is no map $g:\mathbb{C}^d \rightarrow \{0,1,\ldots,d-1\}$ for $d\ge 3$ such that for any pair $(u,v)$ of rays in $\mathbb{C}^d$, $g(u) \neq g(v)$ \cite{Cabello11}. Type (ii) includes all type (i), but there are type (ii) sets which are not type (i) \cite{YO11,BBC11}.

 Recently, Yu and Oh have presented some SIC sets in dimensions $d\ge4$ with a small number of rays \cite{YO12}. The aim of this note is to show that the results in \cite{Cabello11} also allow us to construct even simpler SIC sets. In addition, we clarify which are, so far, the minimum SIC and KS sets in any dimension, identifying those cases in which there is still room for improvement.

Theorem 1 in \cite{Cabello11} states that a set $C \in \mathbb{C}^d$ proves SIC if and only if the graph in which the vertices are all the vectors in $C$ and edges link vectors if and only if they are orthogonal has chromatic number greater than $d$. Theorem 2 in \cite{Cabello11} shows that the smallest SIC set in $d=3$ is the 13-ray set in \cite{YO11}.

{\bf Corollary (Simple SIC sets): }{\em For every $d\ge3$ there is a SIC set with $d+10$ rays.}

{\em Proof: }Let $S_3$ be the the 13-ray set in $d=3$ in \cite{YO11}. Define $S_{3+j}=\{(x,0,\ldots,0) \in \mathbb{C}^{3+j}\;|\;x \in S_3\} \cup \{(0,0,0,1,0,\ldots,0)$, $(0,0,0,0,1,0,\ldots,0),\ldots$, $(0,\ldots,0,1) \in \mathbb{C}^{3+j}\}$, with $j=0,1,2\ldots$ The graph in which the vertices are all the rays in $S_{3+j}$ and edges link rays if and only if they are orthogonal has chromatic number $4+j$.\hfill \endproof

An explicit inequality to test SIC using $d+10$ rays in dimension $d\ge4$ can be constructed using Theorem 3 in \cite{Cabello11}, taking into account that all the observables can be expressed in terms of the rank-1 projectors corresponding to the SIC set.

The previous result improves those presented in \cite{YO12} for any $d\ge 4$, and establishes new records in any of these dimensions. A still open question is which are the minimum SIC sets in any $d\ge 4$. Preliminary results show that, for $d=4$, the minimum SIC set must have between 11 and 14 rays \cite{CP12}.

Table \ref{TableI} indicates the state of the art for minimal SIC and KS sets in dimensions $3\ge d \ge 8$. There are only two sets which have been proven to be minimum: the SIC set in $d=3$ and the KS set in $d=4$. It has also been proven that there are no KS set in $d=3$ with 18 or less rays and that there are no SIC set in $d=4$ with 10 or less rays. In addition, there are explicit SIC and KS sets in all dimensions providing upper bounds.


\begin{table}
\begin{center}
\begin{tabular}{ccc}
\hline
Dimension & Minimum SIC set & Minimum KS set \\
\hline
3 & $13$ \cite{YO11,Cabello11} & $19$ \cite{PMMM05,Cabello06,AOW11} $\le\;? \le 31$ \cite{Peres95}\\
4 & $11$ \cite{CP12} $\le\; ? \le 14$ (This work) & $18$ \cite{CEG96,PMMM05,Cabello06} \\
5 & $? \le 15$ (This work) & $? \le 29$ \cite{CEG05} \\
6 & $? \le 16$ (This work) & $? \le 31$ \cite{CEG05} \\
7 & $? \le 17$ (This work) & $? \le 34$ \cite{CEG05} \\
8 & $? \le 18$ (This work) & $? \le 36$ \cite{KP95,CEG05} \\
\hline
\end{tabular}
\end{center}
\caption{Minimum number of rays needed to prove SIC in dimensions 3--8 with and without KS sets. ``?'' indicates that the actual number is still unknown.}
\label{TableI}
\end{table}


\begin{acknowledgments}
The author thanks I. Bengtsson, K. Blanchfield, J.-{\AA}. Larsson, and J. R. Portillo for discussions. This work was supported by the Projects No.\ FIS2008-05596 and No.\ FIS2011-29400, and the Wenner-Gren Foundation.
\end{acknowledgments}




\begin{thebibliography}{99}


\bibitem{Specker60}
 E. P. Specker,
 Dialectica {\bf 14}, 239 (1960).

\bibitem{Bell66}
 J. S. Bell,
 Rev. Mod. Phys. {\bf 38}, 447 (1966).

\bibitem{KS67}
 S. Kochen and E. P. Specker,
 J. Math. Mech. {\bf 17}, 59 (1967).


\bibitem{Cabello11}
 A. Cabello,
 \eprint{arXiv:1112.5149v1}.


\bibitem{YO11}
 S. Yu and C. H. Oh,
 Phys. Rev. Lett. (in press);
 \eprint{arXiv:1109.4396v1}.

\bibitem{BBC11}
 I. Bengtsson, K. Blanchfield, and A. Cabello,
 Phys. Lett.~A \textbf{376}, 374 (2012).

\bibitem{YO12}
 S. Yu and C. H. Oh,
 \eprint{arXiv:1112.5513v1}.

\bibitem{CP12}
 A. Cabello {\em et al.} (in preparation).


\bibitem{PMMM05}
 M. Pavi\v{c}i\'{c}, J.-P. Merlet, B. D. McKay, and N. D. Megill,
 J. Phys.~A \textbf{38}, 1577 (2005).

\bibitem{Cabello06}
 A. Cabello,
 Int. J. Quant. Info. \textbf{4}, 55 (2006).

\bibitem{AOW11}
 F. Arends, J. Ouaknine, and C. W. Wampler,
 in {\em Graph-Theoretic Concepts in Computer Science},
 edited by P. Kolman and J. Kratochv\'{\i}l,
 Lecture Notes in Computer Science \textbf{6986} (Springer, Berlin, 2011), p.~23.

\bibitem{Peres95}
 A. Peres,
 \emph{Quantum Theory: Concepts and Methods}
 (Kluwer, Dordrecht, 1995), p.~114.

\bibitem{CEG96}
 A. Cabello, J. M. Estebaranz, and G. Garc\'{\i}a-Alcaine,
 Phys. Lett.~A \textbf{212}, 183 (1996).

\bibitem{CEG05}
 A. Cabello, J. M. Estebaranz, and G. Garc\'{\i}a-Alcaine,
 Phys. Lett.~A \textbf{339}, 425 (2005).

\bibitem{KP95}
 M. Kernaghan and A. Peres,
 Phys. Lett.~A {\bf 198}, 1 (1995).


\end{thebibliography}
\end{document}